# Electro-thermal Co-design of Vertical β-$Ga_2O_3$ Schottky Diodes with High-permittivity $BaTiO_3$ Field-plate for High-field and Thermal Management

Ahsanul Mohaimeen Audri,[1] Chung-Ping Ho,[2] Emerson J. Hollar,[1] Jingjing Shi,[2] and Esmat Farzana[1,*]

[1]Department of Electrical and Computer Engineering, Iowa State University, Ames, IA 50011, USA

[2]Department of Mechanical and Aerospace Engineering, University of Florida, Gainesville, FL 32611, USA

**Abstract**— This work presents electrothermal co-design of vertical β-$Ga_2O_3$ Schottky barrier diodes (SBDs) to enhance both heat dissipation and high field management in high-power applications. Here, we demonstrate device-level thermal management tailored for two vertical β-$Ga_2O_3$ SBD structures that employed different edge termination techniques, such as field-plate and deep etch with sidewall field-plate, where the field-plate was formed with high-permittivity dielectric ($BaTiO_3$). The localized thermal hot spots were detected at the Schottky contact edges near $BaTiO_3$ dielectric based field-plate. However, a substantial reduction of the thermal hotspots was observed by forming the field-plate with $BaTiO_3$ and thermally-conductive AlN insulator, where the AlN can effectively decrease Joule heating at interface and the high permittivity of $BaTiO_3$ contributes to high field reduction. The deep etch and sidewall field-plate SBD structure further reduced accumulated heat and electric field near the critical anode edge by removing lateral depletion regions. We also analyzed thermal transport at dielectric/β-$Ga_2O_3$ interfaces using Landauer approach that revealed significantly higher thermal boundary conductance (TBC) enabled by AlN compared to $BaTiO_3$, attesting to the superior heat dissipation ability by $BaTiO_3$/AlN field-plate than the $BaTiO_3$-only configuration. Experimental investigation with vertical metal/AlN/β-$Ga_2O_3$ diodes also extracted a high breakdown field (~11 MV/cm) of AlN, significantly exceeding the material breakdown field of β-$Ga_2O_3$. This indicates that AlN can be an excellent choice for field-plate dielectric in vertical β-$Ga_2O_3$ SBDs to provide both enhanced high field sustainability and improved heat dissipation in high-power applications.

*E-mail: efarzana@iastate.edu

The ultra-wide bandgap (UWBG) semiconductor, β-$Ga_2O_3$, has gained great interest for high-power electronics due to its large bandgap (~4.8eV), predicted high critical breakdown field (8 MV/cm), shallow n-type dopants, and the availability of melt-grown native substrates that enable cost-effective, large-area production of high-quality homoepitaxial films.[1-3] The continued progress in epitaxial growth, device processing, and field management techniques have also led to significant advancement in high-power device performance.[2-12] Among these, vertical n-type Schottky barrier diodes (SBDs) have remained primary interest of research for high-voltage devices due to the absence of p-type β-$Ga_2O_3$.[13] Various advanced field management strategies have been adopted in vertical β-$Ga_2O_3$ devices including field plate,[14,15] guard ring,[16] deep etch,[17] ion-implantation,[18] and trench diodes,[19] that have resulted in kV-class power devices with promising performance metrics.

However, a key requirement for robust high-power devices is to have the ability to manage both extreme heat and high field effects to prevent accelerated electrical or thermal failure. This is particularly challenging for β-$Ga_2O_3$ devices due to the low inherent thermal conductivity of β-$Ga_2O_3$ (11 -24 W/m K), which is an order of magnitude lower than other (U)WBG semiconductors, such as GaN, SiC, and AlN.[20,21] This low thermal conductivity has been reported to cause significant degradation of device performance in β-$Ga_2O_3$ MOSFETs from self-heating effects, such as dramatic thermal droop[22] as well as early catastrophic failure.[23] To circumvent this limitation, efficient β-$Ga_2O_3$ device design strategies need to be developed that can address both high field and thermal effects in high-power applications. However, to date, most β-$Ga_2O_3$ device reports have focused on maximizing the device performance by managing high field effects,[14,16,18,24-30] without incorporating thermal considerations. Although there has been some investigations on thermal management of β-$Ga_2O_3$ devices, the existing reports predominantly focused on lateral device structures including electrothermal modeling,[20,31] top side passivation with heat spreading layers (AlN, nanocrystalline diamond),[32] and heterogenous integration with thermally conductive substates (SiC, Diamond).[33,34] Compared to the lateral counterparts, thermal management of vertical power devices remains relatively sparse. For example, electro-thermal modeling and thermal characterization were performed in bare vertical β-$Ga_2O_3$ Schottky diodes on highly doped β-$Ga_2O_3$ substrates.[35] However, these devices did not include the vertical high-voltage SBD structure that requires low-doped drift layer and field management. Thermal modeling of vertical β-$Ga_2O_3$ trench-Fin MOSFET[36] and Current Aperture Vertical Effect

Transistor (CAVET)[37] were also investigated with Joule-heat power profile mapped at the on-state condition. Moreover, thermal simulation of vertical β-$Ga_2O_3$ SBD with low-permittivity dielectric ($Si_3N_4$) field-plate was reported for different substrate and epilayer thicknesses.[38] However, the off-state performance under high electric field was not investigated for these vertical β-$Ga_2O_3$ devices. [36-38] Hence, a detailed electro-thermal co-design of vertical β-$Ga_2O_3$ high-voltage SBDs needs to be explored since these are the building blocks of multi-kV power switches.

In this work, we performed electrothermal co-design of vertical β-$Ga_2O_3$ SBDs by combining thermal engineering and efficient edge termination strategies that can enable both enhanced heat transport and high field management in high power applications. Towards this goal, we investigated two configurations of vertical β-$Ga_2O_3$ power diode structures, such as basic field-plate (FP) and deep-etch with sidewall field-plate devices. In both structures, we demonstrated that a field-plate dielectric stack of $BaTiO_3$/AlN can simultaneously achieve both high field and thermal management where the top high-permittivity (κ) dielectric $BaTiO_3$ contributes to field reduction due to its extremely high permittivity (~250)[15, 26] and the bottom AlN allows superior heat dissipation owing to its excellent thermal conductivity. The AlN can further improve reverse blocking capability at β-$Ga_2O_3$ interface owing to its higher predicted critical breakdown field (15.4 MV/cm) as well as higher conduction band offset (reported $\Delta E_C$ =0.6 to 1.34 eV) with β-$Ga_2O_3$ compared to that of $BaTiO_3$ ($\Delta E_C$~0.08 eV).[25, 39, 40] To gain insights about the thermal transport, we also investigated thermal boundary conductance (TBC) of AlN and $BaTiO_3$ at β-$Ga_2O_3$ interface using Landauer approach that demonstrated significantly higher TBC obtained with AlN compared to that of $BaTiO_3$. Furthermore, we experimentally evaluated the breakdown field of AlN using Metal-Insulator-Semiconductor (MIS) diodes fabricated on commercially available halide vapor phase epitaxy (HVPE)-grown (001) β-$Ga_2O_3$ samples which showed an excellent electrical breakdown field of AlN (~11 MV/cm), higher than that of β-$Ga_2O_3$ (~8 MV/cm), that will also enhance the breakdown field at the field-plate interface. The AlN also offers growth feasibility using conventional dielectric deposition methods as opposed to diamond that is often challenged by poor seeding density, nucleation difficulties, and substrate degradation during its growth on β-$Ga_2O_3$ [41, 42]. The sputtered AlN has already been demonstrated to enhance top-side heat dissipation in fabricated β-$Ga_2O_3$ and GaN-based lateral transistors[32,43] but has not been utilized yet in vertical β-$Ga_2O_3$ power SBDs. Hence, our work has focused on the efficacy of AlN to achieve device-level thermal management in vertical β-$Ga_2O_3$ power SBDs.

The electrothermal co-design of the vertical β-$Ga_2O_3$ power SBDs was performed using Silvaco ATLAS TCAD that employs a 2-D device model within a self-consistent simulation framework.[44] The dimensions and parameters of the simulated β-$Ga_2O_3$ SBD structures were adopted from the experimental reports of our prior work as well as other published litertaure.[10, 14,15] As most vertical β-$Ga_2O_3$ SBDs utilized homoepitaxial HVPE-grown (001) β-$Ga_2O_3$ thick drift layer on Sn-doped substrate, we also performed electrothermal modeling of vertical SBDs consisting of 10 μm (001) β-$Ga_2O_3$ drift layer (background doping = $1\times10^{16}$ $cm^{-3}$) on a highly-doped ($5\times10^{18}$ $cm^{-3}$) β-$Ga_2O_3$ substrate with substrate thickness of 650 μm.[10] The 100 μm diameter SBDs consisted of Pt/β-$Ga_2O_3$ Schottky contact that enabled a Schottky barrier height (SBH) of 1.5 eV[10, 14] and a backside Ohmic formed with Ti/Au (Fig. 1). The TCAD simulation was conducted with Joule-heat model that included heat capacity, temperature-dependent thermal conductivity ($k_T$), and temperature-dependent mobility ($\mu_T$) based on previous reports as provided in Table I.[37,44-46,49] The electron mobility at 300 K ($\mu_{n,300}$) for the drift layer and the substrate were used 150 and 60, respectively, according to their doping concentration available from experimental results.[47] The temperature-dependent parameters, such as $k_T$ and $\mu_T$, get self-consistently updated in TCAD with the lattice temperature ($T_L$). The TBC of Pt/β-$Ga_2O_3$ Schottky contacts and Ti/β-$Ga_2O_3$ Ohmic contacts were used as 91.4 $MWm^{-2}K^{-1}$ and 103 $MWm^{-2}K^{-1}$, respectively, which were found to be fairly constant across a wide range of temperature (200 -500 K) obtained from the Landauer model as reported in our prior work.[49] The thermal conductivity of AlN was used as 50 $Wm^{-1}K^{-1}$, considering the similar film thickness used in a prior electrothermal modeling of β-$Ga_2O_3$ lateral devices.[31] The Fermi–Dirac model was also used throughout to get accurate results for all doping concentration cases.

To perform the electrothermal TCAD simulation, we implemented the baseline Pt/β-$Ga_2O_3$ vertical Schottky diode with $BaTiO_3$ field plate following the experimentally fabricated device structure reported in Ref [15]. The device had a 25 μm field-plate with a 250 nm $BaTiO_3$ as shown in Fig. (1a).[15] The TCAD simulation, performed at a forward bias of 4 V, showed a concentrated Joule heat hotspot appearing at the Schottky contact edge near field plate (Fig. 2a). However, the peak Joule heat power was reduced by about an order of magnitude, corresponding to an ~ 92% decrease, when the field-plate was formed with a stack dielectric of $BaTiO_3$ (100 nm)/AlN (150 nm) as shown in Fig. 2(b). This demonstrates that the superior thermal conductivity of AlN can facilitate improved heat dissipation and reduce localized thermal hotspot at the field-plate.

Table 1: Key models and parameters used in electrothermal simulations [31,37,45,46,49-52]

| Parameters | Values |
| --- | --- |
| Heat Capacity of β-$Ga_2O_3$ | 3.6 $JK^{-1}cm^{-3}$ |
| TBC at Pt/ β-$Ga_2O_3$ interface | 91.4 $MWm^{-2}K^{-1}$ |
| TBC at Ti/ β-$Ga_2O_3$ interface | 103 $MWm^{-2}K^{-1}$ |
| Thermal Conductivity of β-$Ga_2O_3$ | $k_T=13.7\times(\frac{T_L}{300})^{-1.12}$ $Wm^{-1}K^{-1}$<br>$T_L$= lattice temperature |
| Electron Mobility of β-$Ga_2O_3$ | $\mu_T = \mu_{n,300} \times (\frac{T_L}{300})^{-1.8}$ $cm^2V^{-1}s^{-1}$<br>$\mu_{n,300}$=150 (drift region)/60(substrate) |
| Thermal Conductivity of $BaTiO_3$ | 1.3 $Wm^{-1}K^{-1}$ |
| Heat Capacity of $BaTiO_3$ | 2.61 $JK^{-1}cm^{-3}$ |
| Thermal Conductivity of AlN | 50 $Wm^{-1}K^{-1}$ |
| Heat Capacity of AlN | 2.28 $JK^{-1}cm^{-3}$ |

To investigate the reverse bias performance of these diodes, we performed electric field analysis using TCAD at a reverse bias of 2000 V based on the experimental results of Ref [15] that reported a 2.1 kV breakdown voltage of the fabricated $BaTiO_3$ field-plate SBDs. The peak electric field value near the thermal hotspot regions were obtained as 2.9 $MVcm^{-1}$ and 3.1 $MVcm^{-1}$ in $BaTiO_3$ and $BaTiO_3$/AlN field-plate SBDs, respectively (Fig. 3). The role of AlN is further evident in improving the reverse blocking performance of the device. The peak electric field of the $BaTiO_3$ field-plate diode appears in the β-$Ga_2O_3$ drift layer (Fig. 3a). In contrast, in the $BaTiO_3$/AlN FP diode, the peak field appears in AlN (Fig. 3b) which has a superior critical breakdown field (15.4 $MVcm^{-1}$) than that of β-$Ga_2O_3$ (8 $MVcm^{-1}$). Thus, the electro-thermal co-design with two insulators enables to utilize their complementary properties and achieve both Joule heat reduction and improved reverse blocking performance in vertical β-$Ga_2O_3$ SBDs.

We also investigated the Joule heat power profile as a function of field-plate length ($L_{FP}$) where $L_{FP}$ was varied from 10 μm to 30 μm for both the $BaTiO_3$ and $BaTiO_3$/AlN FP vertical β-$Ga_2O_3$ SBDs (Fig. 2c). It was observed that the Joule heat power density increases with the field-plate length and tends to saturate when $L_{FP}$ increases beyond 20 μm. However, the stack field-plate $BaTiO_3$/AlN SBD consistently provided lower Joule power density by 89-92% compared to that of $BaTiO_3$ for all respective field-plate lengths. It should be noted that although shorter field-plates ($L_{FP}$<20 μm) can enable somewhat reduced Joule-heat power density, such short field-plates would not be sufficient for multi-kV power devices since most experimental vertical β-$Ga_2O_3$ SBDs, with breakdown voltage of 2 kV or higher, had a minimum $L_{FP}$=25 μm where the field-plate was used as the stand-alone edge-termination strategy.[10, 15]

Hence, we investigated additional edge-termination approaches in combination with field-plate that can further assist field-management as well as mitigation of thermal hotspots in the vertical β-$Ga_2O_3$ SBDs. Towards this goal, we employed deep etch epilayer device structure (Fig. 4a) which was demonstrated as an efficient strategy to improve high-voltage performance of vertical β-$Ga_2O_3$ SBDs by reducing lateral depletion regions and thereby, terminating crowded electric field near the Schottky contact edges.[4, 17, 53] Since the Joule heat hotspot is also concentrated near the Schottky contact edges, the deep etch can further assist in Joule heat reduction by removing the thermal hotspot regions. However, as the deep mesa etching can cause plasma damage effects and sidewall leakage, we considered sidewall passivation and field-plate formation in the deep etch vertical SBD structure. The field-plate was formed with $L_{FP}$=3 μm and top dielectric width, $W_{top}$=4 μm, following Ref. 53 where $L_{FP}$ is defined here as the distance from the mesa etch sidewall to the edge of the field-plate as shown in Fig. 4(a). For the field-plate, the deep-etch structure was also investigated with a stack of $BaTiO_3$ (100 nm)/AlN (150 nm) to keep consistency in the thicknesses of the respective dielectric layers as Fig. 1(b). The Joule heat power was evaluated at a forward bias of 4 V for the 100 μm diameter Pt/β-$Ga_2O_3$ SBDs with different etch depths of 0.2, 0.4, 0.6, and 0.8 μm. Figures 4(b) and (c) show the contour plot of the Joule heat power profile for the representative diodes with 0.2 and 0.6 μm deep etch. The Joule heat power at the anode edge and trench corners were further extracted using Cutline AA' for all etch depths as shown in Figs. 4(d) and 4(e). With increased etch depth, a significant reduction of peak Joule heat power density was achieved near anode edges (Fig. 4d), by more than an order of magnitude for all cases, compared to that of the planer $BaTiO_3$/AlN SBD of Fig. 1(b). For these deep etch and side-wall passivated structures, the Joule heat power density got maximized at the trench corners which was also 44-50% lower than that of the planer $BaTiO_3$/AlN SBD structures for all etch depths, and moved further away from the anode edge with increasing etch depth (Fig. 4e). The maximization of Joule heat at trench corners away from the Schottky contact is likely due to removal of lateral depletion regions near anode edge and increased sidewall depletion regions near trench corners.[4,19]

We also evaluated the thermal resistance ($R_{th}$) of the β-$Ga_2O_3$ SBDs with $BaTiO_3$ versus $BaTiO_3$/AlN field-plate. The $R_{th}$ was calculated from the lattice temperature increase with respect to the dissipated power, obtained as [54],

$$R_{th}=\frac{\Delta T}{Pout} \quad \ldots\ldots\ldots.(1)$$

where $\Delta T$ is increase in the peak lattice temperature with respect to ambient temperature, T=300 K and $P_{out}$=dissipated power at V=4 V. There was a monotonic reduction in lattice temperature with increased etch depth from 0.2 μm to 0.8 μm (Figs. 5a-5c), where $BaTiO_3$/AlN field-plate SBDs consistently showed lower lattice temperature than the $BaTiO_3$ ones. This also translated to reduced thermal resistance of $BaTiO_3$/AlN field-plate SBDs than the $BaTiO_3$ counterparts as shown in Fig. 5d.

The electric field at reverse bias of 2000 V was systematically evaluated for the deep etched SBDs. The representative contour profiles for deep etch $BaTiO_3$/AlN SBD are shown with 0.2 and 0.6 μm etch depths in Figs. 6a and 6b, respectively. The electric field near anode edge was further extracted using the horizontal cutline BB' for the different etch depth cases as demonstrated with Fig. 6(c). With increased etch depth, the peak electric field also reduced at anode edges, obtained as 2.9, 1.8, 1.2, and 0.94, for etch depths of 0.2, 0.4, 0.6, and 0.8 μm, respectively (Fig. 6d). These values are substantially lower than those of the baseline planer $BaTiO_3$/AlN FP SBD structure that demonstrated a peak field at anode edge of 3.1 MV/cm (Fig. 3b). Moreover, for the deep etched $BaTiO_3$/AlN FP SBD, the peak electric field appears in the AlN layer at the trench corners, which has higher critical electric field than β-$Ga_2O_3$. Thus, the $BaTiO_3$/AlN field-plate also offers superior efficiency in field management in the deep etched structures.

To gain insights about the heat dissipation at the field-pate interface, we also investigated TBC at AlN/β-$Ga_2O_3$ and $BaTiO_3$/β-$Ga_2O_3$ interfaces using the Landauer approach. The general form of Landauer formula to calculate the TBC, indicated by $G$, at 3D/3D interface can be expressed as:

$$G=\frac{q}{A\Delta T}=\sum_p \int_0^\infty \frac{1}{4} D_1(\omega) v_1(\omega) \tau_{12}(\omega) \hbar\omega \frac{df_{BE}}{dT} d\omega, \quad (2)$$

where $q$ is the net heat flow rate, $A$ is the cross-sectional area of the interface, T is the temperature, $D$ is the phonon density of states, $f_{BE}$ is the Bose-Einstein distribution function, $\hbar$ is the reduced Planck constant, $\omega$ is the phonon angular frequency, $v$ is the phonon group velocity, $\tau_{12}$ is the transmission coefficient from material 1 to 2 (here from β-$Ga_2O_3$ to $BaTiO_3$ or AlN), and the sum is over all phonon branches.

The phonon dispersion curves of $BaTiO_3$, β-$Ga_2O_3$, and AlN from our prior work and previous literature are shown in Fig. 7(a).[55,56] Using β-$Ga_2O_3$ as material 1 in Equation (2), all the terms in the integral form can be directly decided from the phonon dispersion relation of β-$Ga_2O_3$ while the transmission becomes the only factor to affect $q$ of different materials on the other side of the interface ($BaTiO_3$ or AlN). Using the diffuse mismatch model (DMM),[57] the transmission coefficient is given by:

$$\tau_{12}(\omega) = \frac{\sum_p D_2(\omega)v_2(\omega)}{\sum_p D_1(\omega)v_1(\omega)+\sum_p D_2(\omega)v_2(\omega)}. \quad (3)$$

The calculated transmission coefficients from DMM are shown in Fig. 7(b). With the elastic transmission assumption from DMM, the transmission coefficient from material 1 to 2 at a given frequency is zero when there is no existing phonon mode in material 2, or the phonon group velocity of material 2 is zero at the frequency. Figure 7(b) also shows that the non-zero transmission of β-$Ga_2O_3$ to AlN spans over a wider range of frequency than β-$Ga_2O_3$ to $BaTiO_3$ since in some frequency ranges (e.g. 10-13 THz), there are no possible phonon modes in $BaTiO_3$ corresponding to those in β-$Ga_2O_3$. As a result, there are less phonons in β-$Ga_2O_3$ that can be transmitted through the β-$Ga_2O_3$/$BaTiO_3$ interface compared to β-$Ga_2O_3$/AlN and leads to a lower TBC. The temperature dependence of TBC is shown in Fig. 8. At room temperature, the TBC of AlN/β-$Ga_2O_3$ is 483.5 MW/$m^2$K, significantly higher than that of $BaTiO_3$/β-$Ga_2O_3$ interface with 180 MW/$m^2$K, with similar trend existing to higher temperature. These modeling results theoretically demonstrate that adding an AlN layer can enhance the TBC and improve thermal management.

We also experimentally investigated the electrical breakdown characteristics of AlN since dielectric with high breakdown field is essential for β-$Ga_2O_3$ power devices to mitigate premature device failure from early dielectric breakdown. Hence, we characterized the electrical breakdown field of AlN formed by atomic layer deposition (ALD) in vertical metal/AlN/β-$Ga_2O_3$ MIS diode structures. The MIS diodes were fabricated on HVPE-grown (001) β-$Ga_2O_3$ with ~ 10 μm thick epilayer (doping ~1.5-2.0×$10^{16}$ $cm^{-3}$) on a Sn-doped β-$Ga_2O_3$ substrate. We also compared the breakdown properties of the ALD AlN with ALD $Al_2O_3$ of similar thickness. The ALD $Al_2O_3$ has been widely used to form the interfacing layer in field-plate (high-κ dielectric/$Al_2O_3$) that can protect the device surface from plasma damage effects during subsequent high-κ layer deposition by sputtering.[10, 58] The MIS diodes had backside Ohmic contact formed with Ti (30 nm)/Au (150 nm) which was deposited by e-beam evaporation after backside etching via $BCl_3$ reactive ion etching (RIE). The samples were later subjected to rapid thermal annealing in nitrogen ambient at 470 °C for 1 minute. Subsequently, the $Al_2O_3$ and AlN dielectric layers were deposited by ALD at 300 °C with comparable thickness, ~20 nm and ~18 nm, respectively. The $Al_2O_3$ was deposited via thermal ALD whereas the AlN was deposited using plasma-enhanced ALD (PEALD) under nitrogen plasma at 50 W and 5 mTorr, with an $NH_3$: $N_2$ flow ratio of 30:10 sccm. Finally, the circular anode contacts of 100 μm diameter were formed using e-beam deposited metal stack of Ni (150 nm)/Au (30 nm) (Fig. 9).

After the MIS diode fabrication, the forward breakdown characteristics were investigated to ensure that the voltage is almost entirely held in the dielectric layers.[59] As shown in Fig. 9 (c), the AlN MIS diodes exhibited minimal leakage and significantly higher breakdown field of ~ 11 MV/cm compared to that of $Al_2O_3$ (~6 MV/cm). With a breakdown field exceeding the theoretical breakdown field β-$Ga_2O_3$ (8 MV/cm), the AlN can be a preferred choice for the field-plate dielectric stack to support enhanced breakdown capability at the field-plate interface compared to the conventional $Al_2O_3$ that exhibited a breakdown field lower than that of β-$Ga_2O_3$.

In summary, we demonstrated electrothermal co-design of vertical β-$Ga_2O_3$ Schottky diodes combining both material properties and device architectures that can provide excellent field termination as well as device-level thermal management to adapt to the challenges of low thermal conductivity of β-$Ga_2O_3$. The Joule heat modeling revealed a thermal hotspot at the Schottky contact edge near extreme-κ dielectric $BaTiO_3$ field-plate. However, the field-plate design with a stack of $BaTiO_3$ and interface layer of AlN was found to substantially reduce the localized heat accumulation in the baseline field-plate structures. Further improvement in heat dissipation was achieved with the deep etch and sidewall $BaTiO_3$/AlN field-plate SBD structure that enables reduction of the accumulated heat near anode edges by eliminating lateral depletion regions.

Thermal transport analysis at dielectric/β-$Ga_2O_3$ interfaces demonstrated a significantly higher thermal boundary conductance enabled by AlN compared to that of $BaTiO_3$, consistent with the observed superior heat dissipation by $BaTiO_3$/AlN field-plate than the $BaTiO_3$-only counterpart. Moreover, improved reverse blocking capabilities were obtained by $BaTiO_3$/AlN field-plate in both baseline field-plate and deep etch diodes, owing to the high breakdown field of AlN. Our experimental results with vertical β-$Ga_2O_3$ MIS diodes also demonstrated a high breakdown field of ~11 MV/cm achieved with ALD AlN, sufficiently exceeding that of conventional ALD $Al_2O_3$ as well as the intrinsic breakdown field of β-$GaO_3$. The combination of high breakdown field and superior thermal boundary conductance of AlN enabled by $BaTiO_3$/AlN field-plate reveals a promising strategy of improving high field and thermal management of vertical β-$Ga_2O_3$ Schottky diodes. Future studies on electrical and thermal characterization of the fabricated devices as well as further improvement of the AlN quality to achieve its theoretical breakdown limit (~15.4 MV/cm) can enable more insights on integrating $BaTiO_3$/AlN to enhance the electrothermal performance of the vertical β-$Ga_2O_3$ Schottky diodes.

**Acknowledgments**

The work at Iowa State University was supported in part by NSF ECCS program (Award No. 2401579) and 2024 ORAU Ralph E. Powe Junior Faculty Enhancement Award. A portion of this work was done in the UCSB Nanofabrication Facility, an open-access laboratory.

**Author declarations**

The authors have no conflicts to disclose.

**Data Availability**

The data that support the findings of this study are available from the corresponding author upon reasonable request.

**References**

1. M. Higashiwaki, K. Sasaki, A. Kuramata, T. Masui, and S. Yamakoshi, *Appl. Phys. Lett.* **100**, 013504 (2012).

2. E. Farzana, A. Bhattacharyya, N. S. Hendricks, T. Itoh, S. Krishnamoorthy, and J. S. Speck, *APL Mater.* **10**, 111104 (2022).

3. W. Li, Z. Hu, K. Nomoto, Z. Zhang, J.-Y. Hsu, Q. T. Thieu, K. Sasaki, A. Kuramata, D. Jena, and H. G. Xing, *Appl. Phys. Lett.* **113**, 202101 (2018).

4. Z. Han, G. Jian, X. Zhou, Q. He, W. Hao, J. Liu, B. Li, H. Huang, Q. Li, X. Zhao, G. Xu, and S. Long, *IEEE Electron Device Lett.* **44**, 1680 (2023).

5. Z. Jian, C. J. Clymore, K. Sun, U. Mishra, and E. Ahmadi, *Appl. Phys. Lett.* **120**, 142101 (2022).

6. J. Yang, S. Ahn, F. Ren, S. J. Pearton, S. Jang, and A. Kuramata, *IEEE Electron Device Lett.* **38**, 906 (2017).

7. P. H. Carey IV, J. Yang, F. Ren, R. Sharma, M. Law, and S. J. Pearton, *ECS J. Solid State Sci. Technol.* **8**, Q3221 (2019).

8. C. Joishi, S. Rafique, Z. Xia, L. Han, S. Krishnamoorthy, Y. Zhang, S. Lodha, H. Zhao, and S. Rajan, *Appl. Phys. Express* **11**, 031101 (2018).

9. K. Sasaki, A. Kuramata, T. Masui, E. G. Víllora, K. Shimamura, and S. Yamakoshi, *Appl. Phys. Express* **5**, 035502 (2012).

10. E. Farzana, S. Roy, N. S. Hendricks, S. Krishnamoorthy, and J. S. Speck, *Appl. Phys. Lett*. **123**, 192102 (2023).

11. J.-S. Li, C.-C. Chiang, X. Xia, T. J. Yoo, F. Ren, H. Kim, and S. J. Pearton, *Appl. Phys. Lett.* **121**, 042105 (2022).

12. Y. Zhang, A. Mauze, F. Alema, A. Osinsky, T. Itoh, and J. S. Speck, *Jpn. J. Appl. Phys.* **60**, 014001 (2021).

13. W. Li, D. Saraswat, Y. Long, K. Nomoto, D. Jena, and H. G. Xing, *Appl. Phys. Lett.* **116**, 192101 (2020).

14. K. Konishi, K. Goto, H. Murakami, Y. Kumagai, A. Kuramata, S. Yamakoshi, and M. Higashiwaki, *Appl. Phys. Lett.* **110**, 103506 (2017).

15. S. Roy, A. Bhattacharyya, C. Peterson, and S. Krishnamoorthy, *Appl. Phys. Lett.* **122**, 152101 (2023).

16. S. Roy, A. Bhattacharyya, and S. Krishnamoorthy, *IEEE Trans. Electron Devices* **67**, 4842 (2020).

17. S. Dhara, N. K. Kalarickal, A. Dheenan, C. Joishi, and S. Rajan, *Appl. Phys. Lett.* **121**, 203501 (2022).

18. C. Lin, Y. Yuda, M. H. Wong, M. Sato, N. Takekawa, K. Konishi, T. Watahiki, M. Yamamuka, H. Murakami, Y. Kumagai, and M. Higashiwaki, *IEEE Electron Device Lett.* **40**, 1487 (2019).

19. W. Li, K. Nomoto, Z. Hu, D. Jena, and H. G. Xing, *Appl. Phys. Express* **12**, 061007 (2019).

20. C. Yuan, Y. Zhang, R. Montgomery, S. Kim, J. Shi, A. Mauze, T. Itoh, J. S. Speck, and S. Graham, *J. Appl. Phys.* **127**, 154502 (2020).

21. Z. Guo, A. Verma, X. Wu, F. Sun, A. Hickman, T. Masui, A. Kuramata, M. Higashiwaki, D. Jena, and T. Luo, *Appl. Phys. Lett.* **106**, 111909 (2015).

22. M. Singh, M. A. Casbon, M. J. Uren, J. W. Pomeroy, S. Dalcanale, S. Karboyan, P. J. Tasker, M. H. Wong, K. Sasaki, A. Kuramata, S. Yamakoshi, M. Higashiwaki, and M. Kuball, *IEEE Electron Device Lett*. **39**, 1572 (2018).

23. A. J. Green, K. D. Chabak, M. Baldini, N. Moser, R. Gilbert, R. C. Fitch Jr., G. Wagner, Z. Galazka, J. McCandless, A. Crespo, K. Leedy, and G. H. Jessen Sr., *IEEE Electron Device Lett.* **38**, 790 (2017).

24. W. Li, K. Nomoto, Z. Hu, D. Jena, and H. G. Xing, *IEEE Electron Device Lett*. **41**, 107 (2020).

25. Z. Xia, H. Chandrasekar, W. Moore, C. Wang, A. J. Lee, J. McGlone, N. K. Kalarickal, A. Arehart, S. Ringel, F. Yang, and S. Rajan, *Appl. Phys. Lett.* **115**, 252104 (2019).

26. S. Roy, A. Bhattacharyya, P. Ranga, H. Splawn, J. Leach, and S. Krishnamoorthy, *IEEE Electron Device Lett*. **42**, 1140 (2021).

27. H. H. Gong, X. X. Yu, Y. Xu, X. H. Chen, Y. Kuang, Y. J. Lv, Y. Yang, F.-F. Ren, Z. H. Feng, S. L. Gu, Y. D. Zheng, R. Zhang, and J. D. Ye, *Appl. Phys. Lett.* **118**, 202102 (2021).

28. Q. Yan, H. H. Gong, J. Zhang, J. Ye, H. Zhou, Z. Liu, S. Xu, C. Wang, Z. Hu, Q. Feng, J. Ning, C. Zhang, P. Ma, R. Zhang, and Y. Hao, *Appl. Phys. Lett.* **118**, 122102 (2021).

29. H. Zhou, Q. Yan, J. Zhang, Y. Lv, Z. Liu, Y. Zhang, K. Dang, P. Dong, Z. Feng, Q. Feng, J. Ning, C. Zhang, P. Ma, and Y. Hao, *IEEE Electron Device Lett.* **40**, 1788 (2019).

30. J. Baliga, *Fundamentals of Power Semiconductor Devices* (Springer, New York, NY, 2008).

31. N. Kumar, C. Joishi, Z. Xia, S. Rajan, and S. Kumar, *Proc. IEEE ITherm*, 370 (2019). doi: 10.1109/ITHERM.2019.8757413.

32. J. S. Lundh, H. N. Masten, K. Sasaki, A. G. Jacobs, Z. Cheng, J. Spencer, L. Chen, J. Gallagher, A. D. Koehler, K. Konishi, S. Graham, A. Kuramata, K. D. Hobart, and M. J. Tadjer, *Proc. Device Res. Conf. (DRC)*, **1** (2022).

33. Y. Song, A. Bhattacharyya, A. Karim, D. Shoemaker, H.-L. Huang, S. Roy, C. McGray, J. H. Leach, J. Hwang, S. Krishnamoorthy, and S. Choi, *ACS Appl. Mater. & Interfaces* **15**, 7137 (2023).

34. X. Chen, F. N. Donmezer, S. Kumar, and S. Graham, *IEEE Trans. Electron Devices* **61**, 4056 (2014).

35. B. Chatterjee, A. Jayawardena, E. Heller, D. W. Snyder, S. Dhar, and S. Choi, *Rev. Sci. Instrum.* **89**, 114903 (2018).

36. R. H. Montgomery, Y. Zhang, C. Yuan, S. Kim, J. Shi, T. Itoh, A. Mauze, S. Kumar, J. Speck, and S. Graham, *J. Appl. Phys.* **129**, 085301 (2021).

37. S. Kim, Y. Zhang, C. Yuan, R. Montgomery, A. Mauze, J. Shi, E. Farzana, J. S. Speck, and S. Graham, *IEEE Trans. Compon. Packag. Manuf. Technol.* **11**, 1171 (2021).

38. R. Sharma, E. Patrick, M. E. Law, J. Yang, F. Ren, and S. J. Pearton, *ECS J. Solid State Sci. Technol.* **8**, Q3195 (2019).

39. J.-X. Chen, J.-J. Tao, H.-P. Ma, H. Zhang, J.-J. Feng, W.-J. Liu, C. Xia, H.-L. Lu, D. W. Zhang, *Appl. Phys. Lett*. **112**, 261602 (2018).

40. K. D. Chabak, K. D. Leedy, A. J. Green, S. Mou, A. T. Neal, T. Asel, E. R. Heller, N. S. Hendricks, K. Liddy, A. Crespo, N. C. Miller, M. T. Lindquist, N. A. Moser, R. C. Fitch Jr., D. E. Walker Jr., D. L. Dorsey, and G. H. Jessen, *Semicond. Sci. Technol.* **35**, 013002 (2020).

41. S. Mandal, K. Arts, H.C.M. Knoops, J.A. Cuenca, G.M. Klemencic, and O.A. Williams, *Carbon* **181**, 296 (2021).

42. M. Malakoutian, Y. Song, C. Yuan, C. Ren, J. Spencer Lundh, R. M. Lavelle, J. E. Brown, D. W. Snyder, S. Graham, S. Choi, and S. Chowdhury, *Appl. Phys. Express* **14**, 055502 (2021).

43. N. Tsurumi, H. Ueno, T. Murata, H. Ishida, Y. Uemoto, T. Ueda, K. Inoue, and T. Tanaka, *IEEE Trans. Electron Devices* **57**, 980 (2010).

44. Atlas User's Manual, atlas_umv1.book

45. Z. Galazka, K. Irmscher, R. Uecker, R. Bertram, M. Pietsch, A. Kwasniewski, M. Naumann, T. Schulz, R. Schewski, D. Klimm, and M. Bickermann, *J. Cryst. Growth* **404**, 184 (2014).

46. M. Xiao, B. Wang, J. Liu, R. Zhang, Z. Zhang, C. Ding, S. Lu, K. Sasaki, G.-Q. Lu, C. Buttay, and Y. Zhang, *IEEE Trans. Power Electron*. **36**, 8565 (2021).

47. Z. Feng, A. F. M. A. U. Bhuiyan, M. R. Karim, and H. Zhao, *Appl. Phys. Lett.* **114**, 250601 (2019).

48. H. Y. Wong, *ECS J. Solid State Sci. Technol*. **12**, 055002 (2023).

49. J. Shi, C. Yuan, H.-L. Huang, J. Johnson, C. Chae, S. Wang, R. Hanus, S. Kim, Z. Cheng, J. Hwang, and S. Graham, *ACS Appl. Mater. Interfaces* **13**, 29083 (2021).

50. J. Cho, J. Park, F. B. Prinz, and J. An, *Scr. Mater*. **154**, 225 (2018).

51. Y. He, *Thermochim. Acta* **419**, 135 (2004).

52. K. Ait Aissa, N. Semmar, A. Achour, Q. Simon, A. Petit, J. Camus, C. Boulmer-Leborgne, and M. A. Djouadi, *J. Phys. D: Appl. Phys.* **47**, 355303 (2014).

53. J. Wan, H. Wang, C. Wang, H. Chen, C. Zhang, L. Zhang, Y. Li, and K. Sheng, *IEEE Electron Device Lett.* **45**, 778 (2024).

54. Y. Dong, E. Yagyu, T. Matsuda, K. H. Teo, C. Lin, and S. Rakheja, *IEEE J. Electron Devices Soc*. **13**, 54 (2025).

55. Z. Cheng, Y. R. Koh, A. Mamun, J. Shi, T. Bai, K. Huynh, L. Yates, Z. Liu, R. Li, E. Lee, M. E. Liao, Y. Wang, H. M. Yu, M. Kushimoto, T. Luo, M. S. Goorsky, P. E. Hopkins, H. Amano, A. Khan, and S. Graham, *Phys. Rev. Mater.* **4**, 044602 (2020).

56. G. Petretto, S. Dwaraknath, H. P. C. Miranda, D. Winston, M. Giantomassi, M. J. van Setten, X. Gonze, K. A. Persson, G. Hautier, and G.-M. Rignanese, *Sci. Data* **5**, 180065 (2018).

57. E. T. Swartz and R. O. Pohl, *Rev. Mod. Phys.* **61**, 605 (1989).

58. N. K. Kalarickal, Z. Xia, H.-L. Huang, W. Moore, Y. Liu, M. Brenner, J. Hwang, and S. Rajan, *IEEE Electron Device Lett*. **42**, 899 (2021).

59. X. Zhai, Z. Wen, O. Odabasi, E. Achamyeleh, K. Sun, and E. Ahmadi, *Appl. Phys. Lett.* **124**, 132103 (2024).

**Figures**

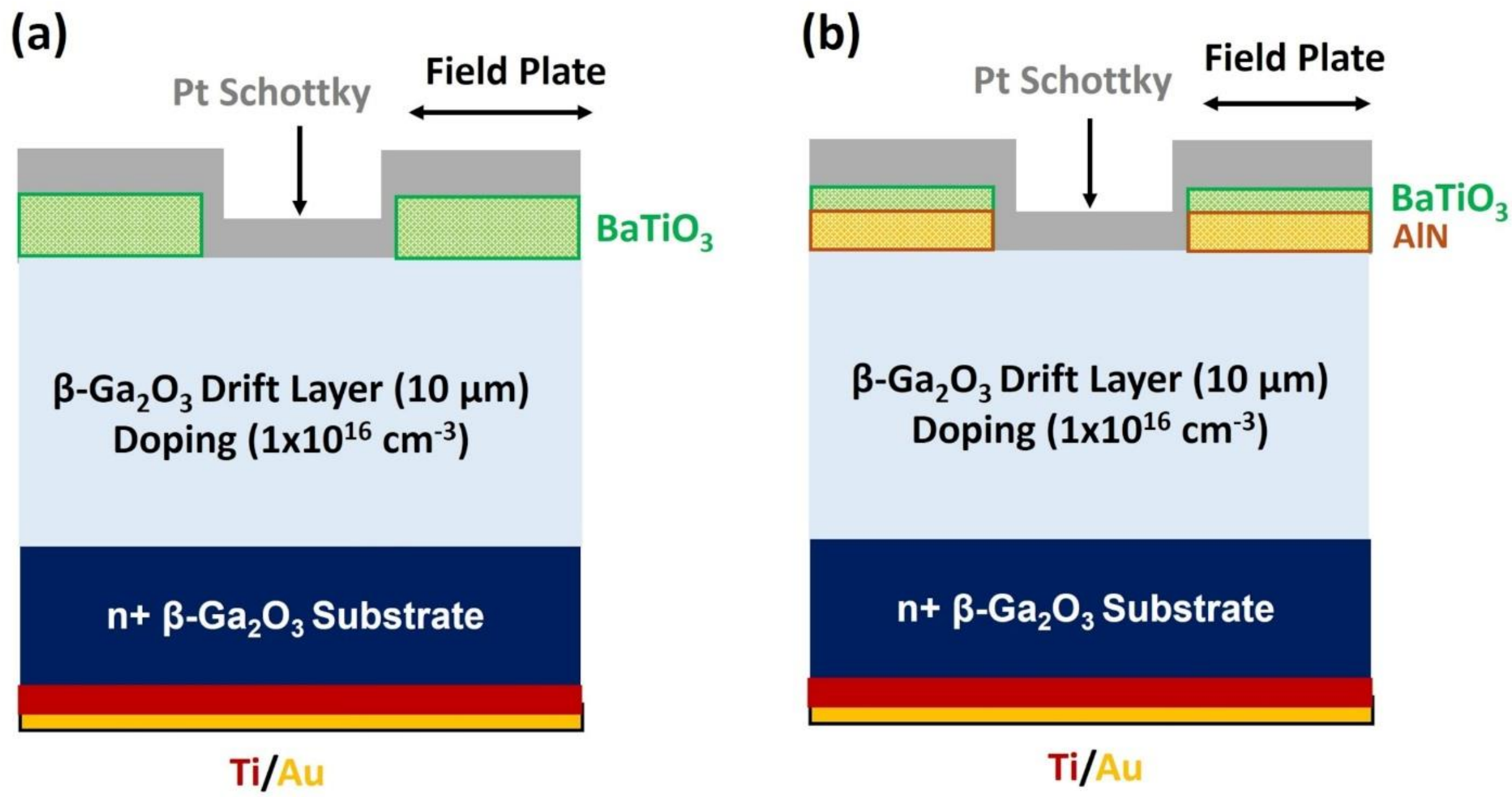

Figure 1. Device schematic of vertical β-$Ga_2O_3$ SBD of 100 µm diameter with (a) $BaTiO_3$ FP, and (b) $BaTiO_3$/AlN FP stack.

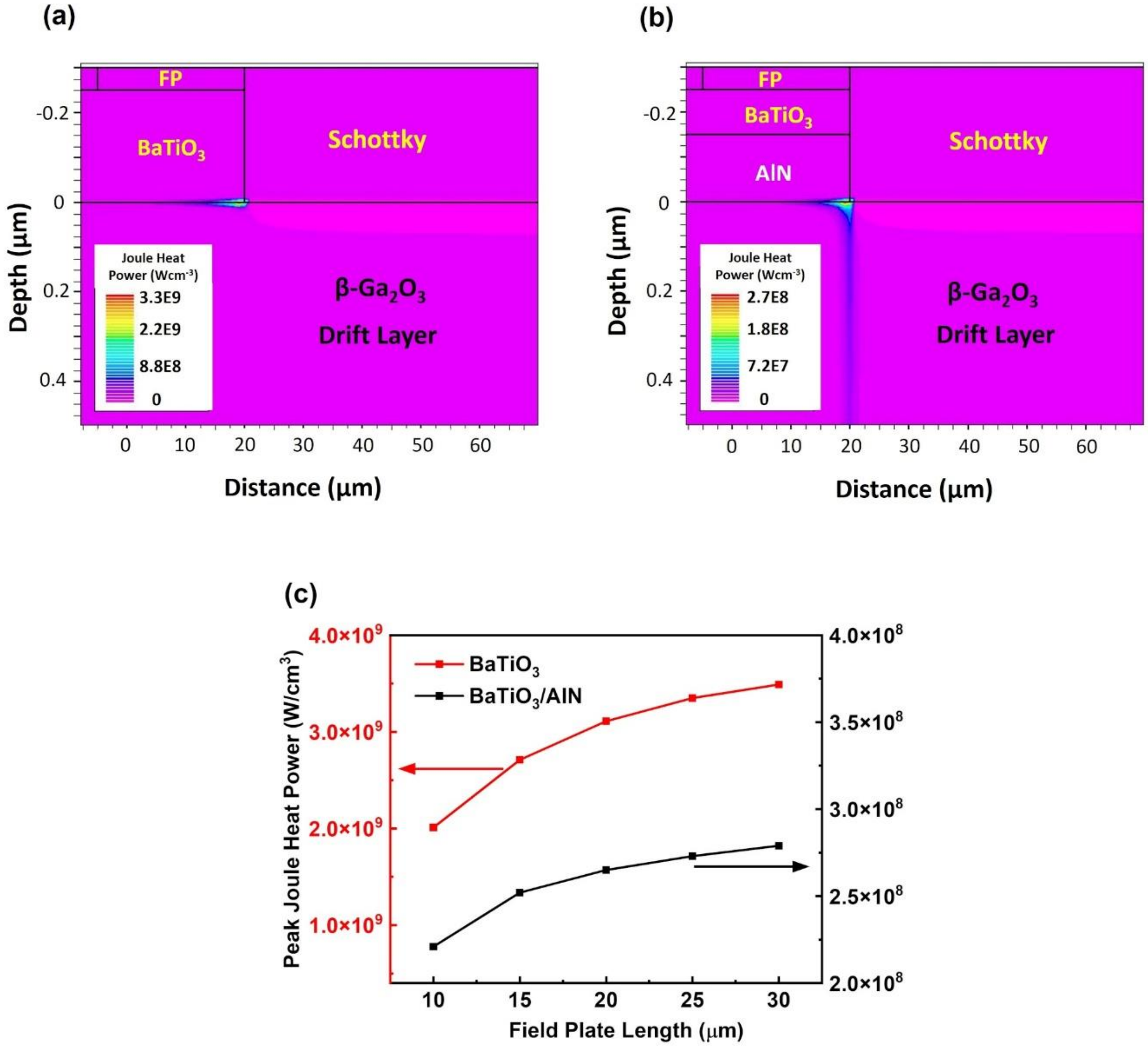

Figure 2. Simulated Joule heat power contour profile of vertical β-$Ga_2O_3$ SBDs at forward bias 4V with (a) $BaTiO_3$ FP and (b) $BaTiO_3$/AlN FP. The $BaTiO_3$/AlN FP provides an order of magnitude reduction of Joule heat near anode edges. (c) The Joule heat power as a function of field-plate length for $BaTiO_3$ (250 nm) and $BaTiO_3$ (100 nm)/AlN (150 nm) FP stack.

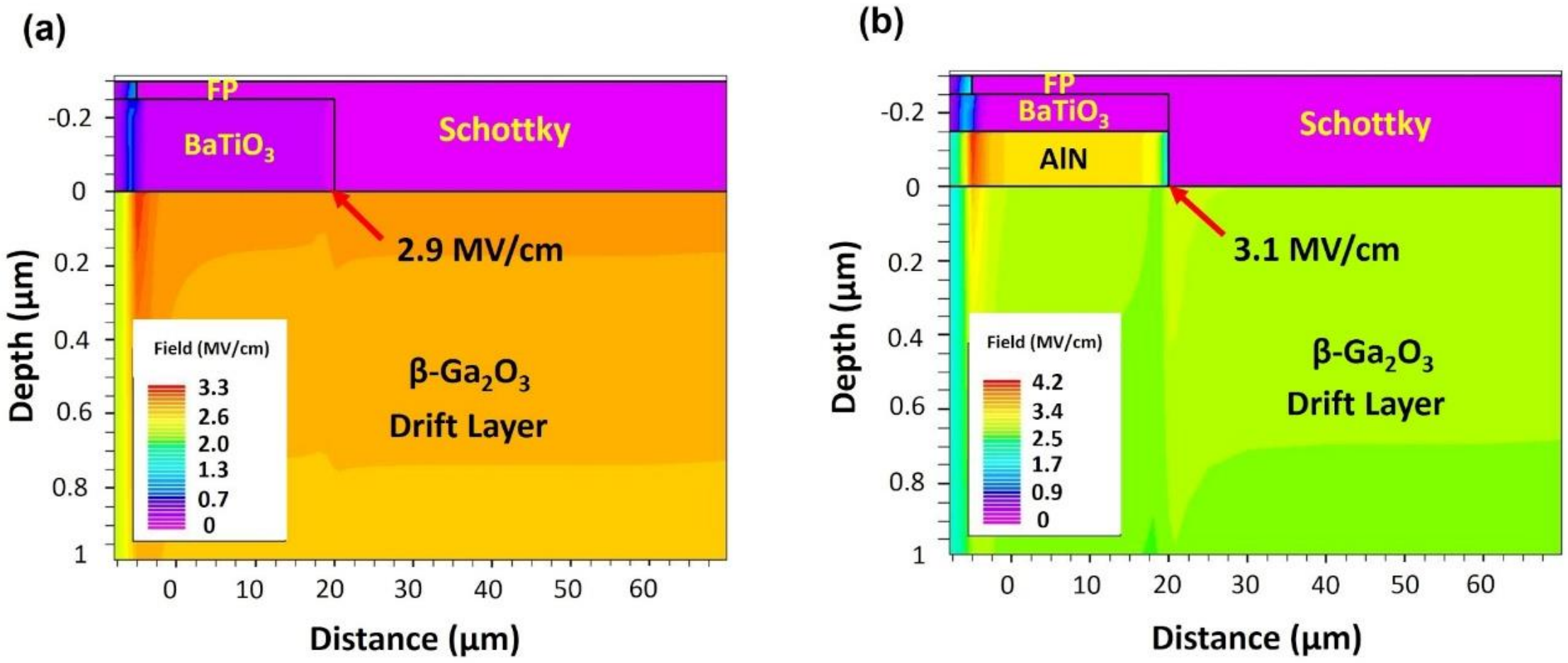

Figure 3. Simulated electric field contour profile of vertical β-$Ga_2O_3$ SBDs at reverse bias 2000 V with (a) $BaTiO_3$ FP and (b) $BaTiO_3$/AlN FP. The peak electric field near the thermal hotspot regions is obtained as 2.9 MV/cm and 3.1 MV/cm in $BaTiO_3$ and $BaTiO_3$/AlN FP SBD, respectively.

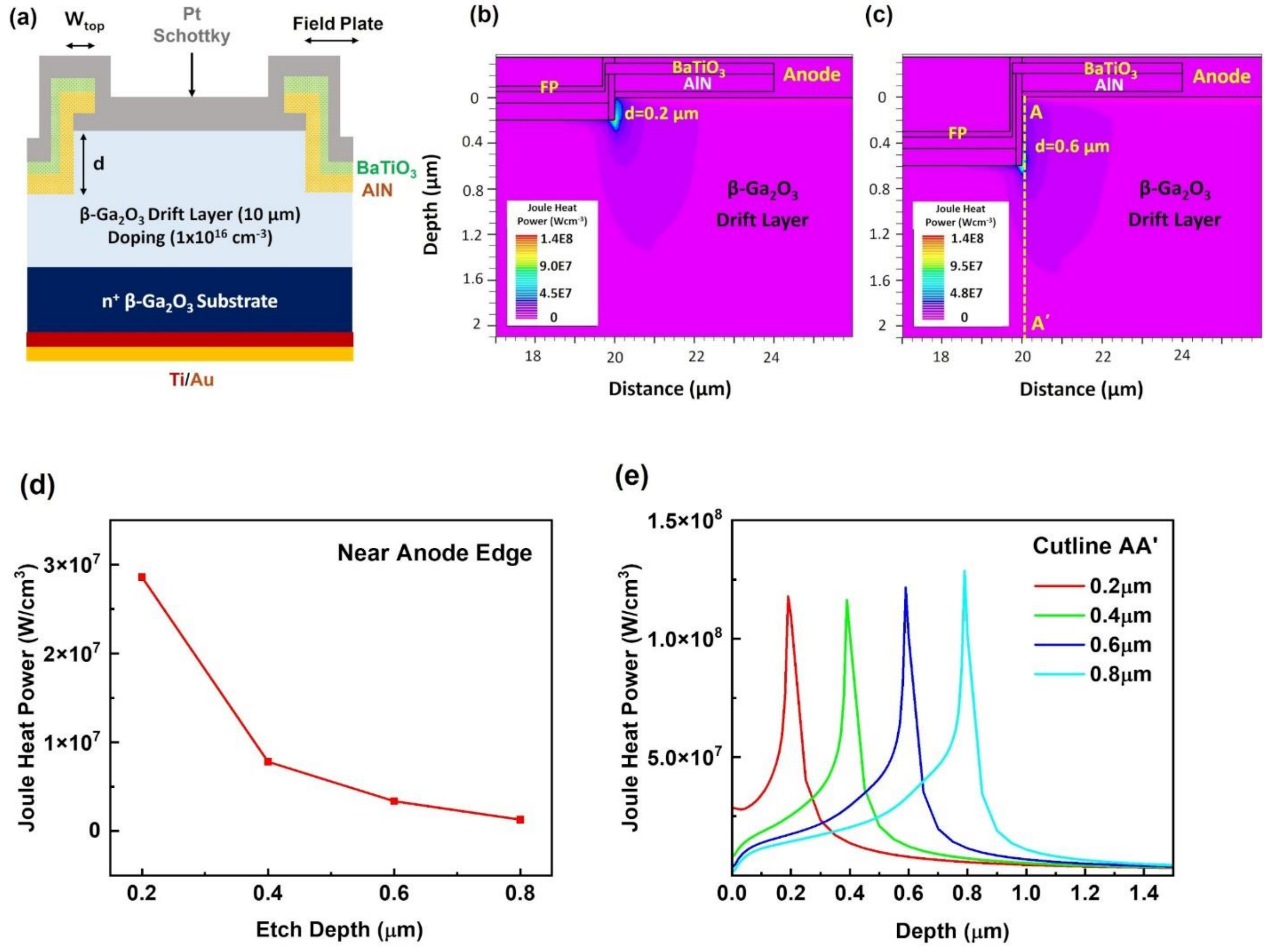

Figure 4. (a) Device schematic of vertical β-$Ga_2O_3$ SBD of 100 µm diameter with deep etch depth (d) and sidewall FP formed with $BaTiO_3$/AlN dielectric stack. Simulated Joule heat power contour profile at forward bias of 4V shown for etch depths of (b) 0.2 µm and (c) 0.6 µm. The cutline AA' was used to extract Joule heat power for different etch depths from 0.2 µm to 0.8 µm at (d) near anode edge indicated by point A of the cutline AA', (e) near trench corner showing the peak Joule heat power appears at the trench corner.

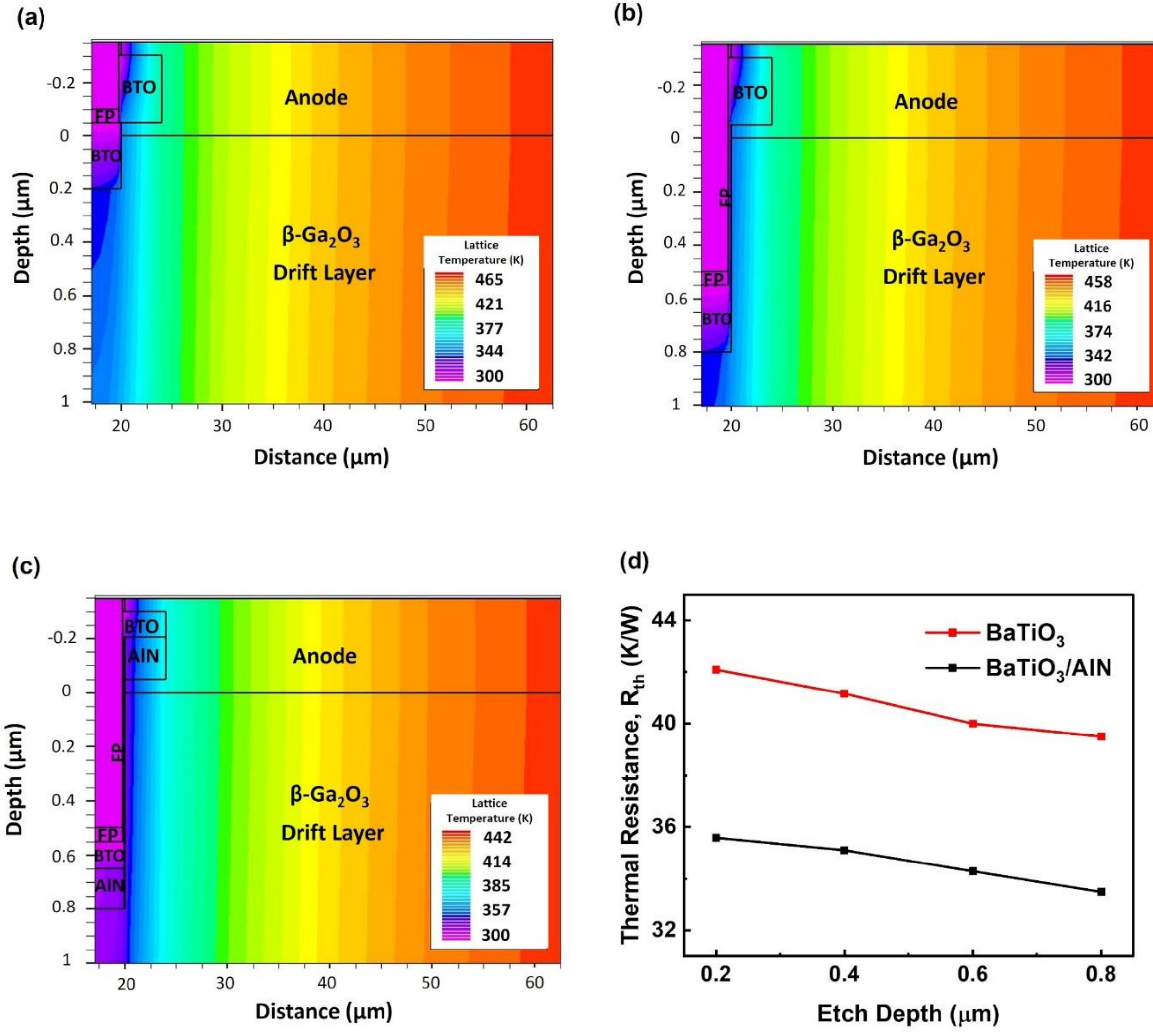

Figure 5. Simulated contour plot of lattice temperature at forward bias of 4V shown for the deep etch vertical β-$Ga_2O_3$ SBDs with etch depths of (a) 0.2 μm with $BaTiO_3$ field-plate, (a) 0.8 μm with $BaTiO_3$ field-plate and (c) 0.8 μm with $BaTiO_3$/AlN stack field-plate. (d) Thermal resistance ($R_{th}$) vs. etch depth for $BaTiO_3$ and $BaTiO_3$/AlN stack field-plate SBDs. The $BaTiO_3$/AlN field-plate SBDs show reduced thermal resistance compared to the $BaTiO_3$ field-plate ones.

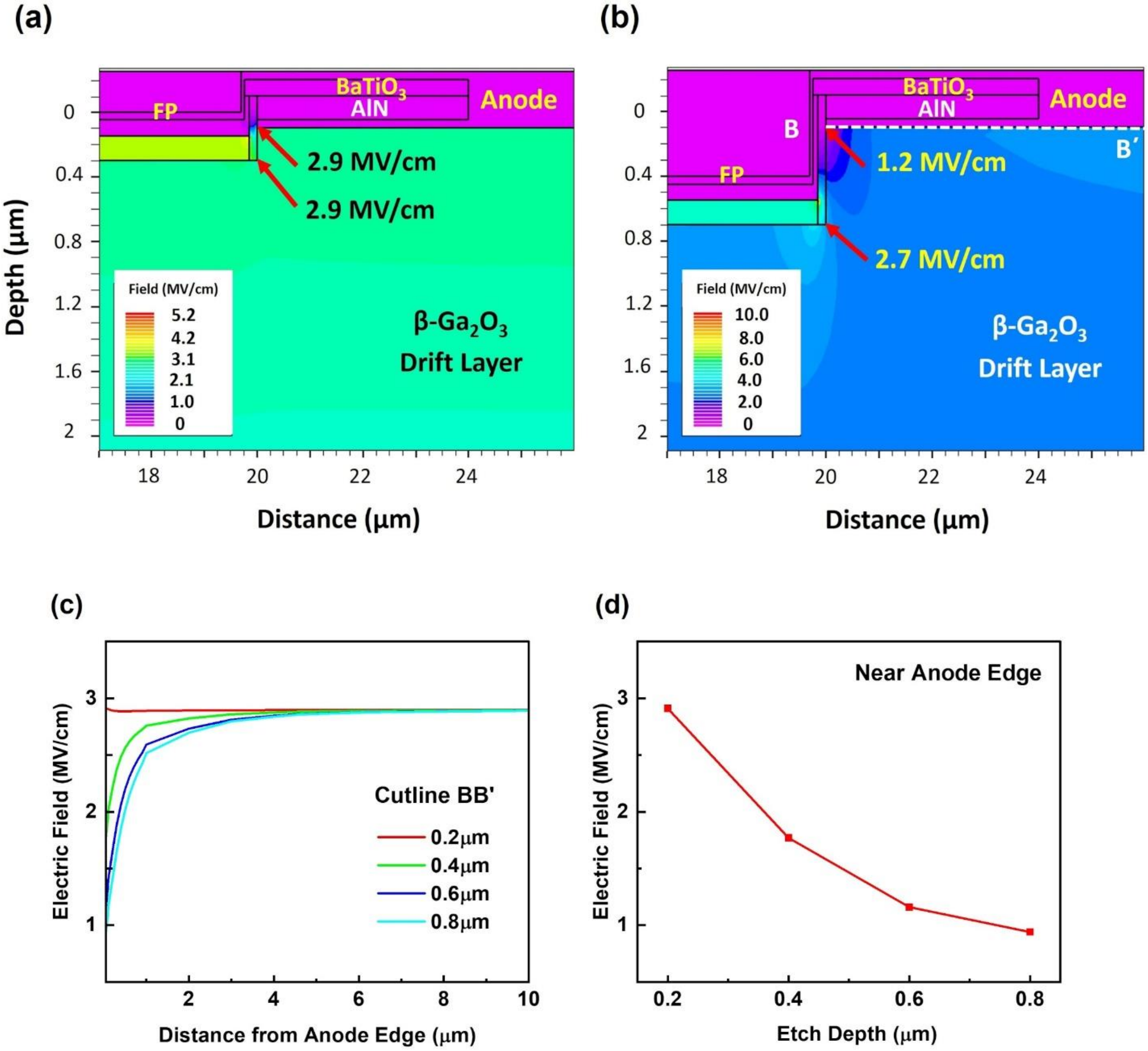

Figure 6. Simulated electric field contour profiles at reverse bias 2000V for the deep etch vertical $\beta$-$Ga_2O_3$ SBDs with $BaTiO_3$/AlN FP shown for etch depths of (a) 0.2 μm and (b) 0.6 μm. (c) The electric field profile along the anode edge extracted using horizontal cutline BB' for different etch depths. (d) Electric field near anode edge as a function of etch depth indicated by point B of the cutline BB'.

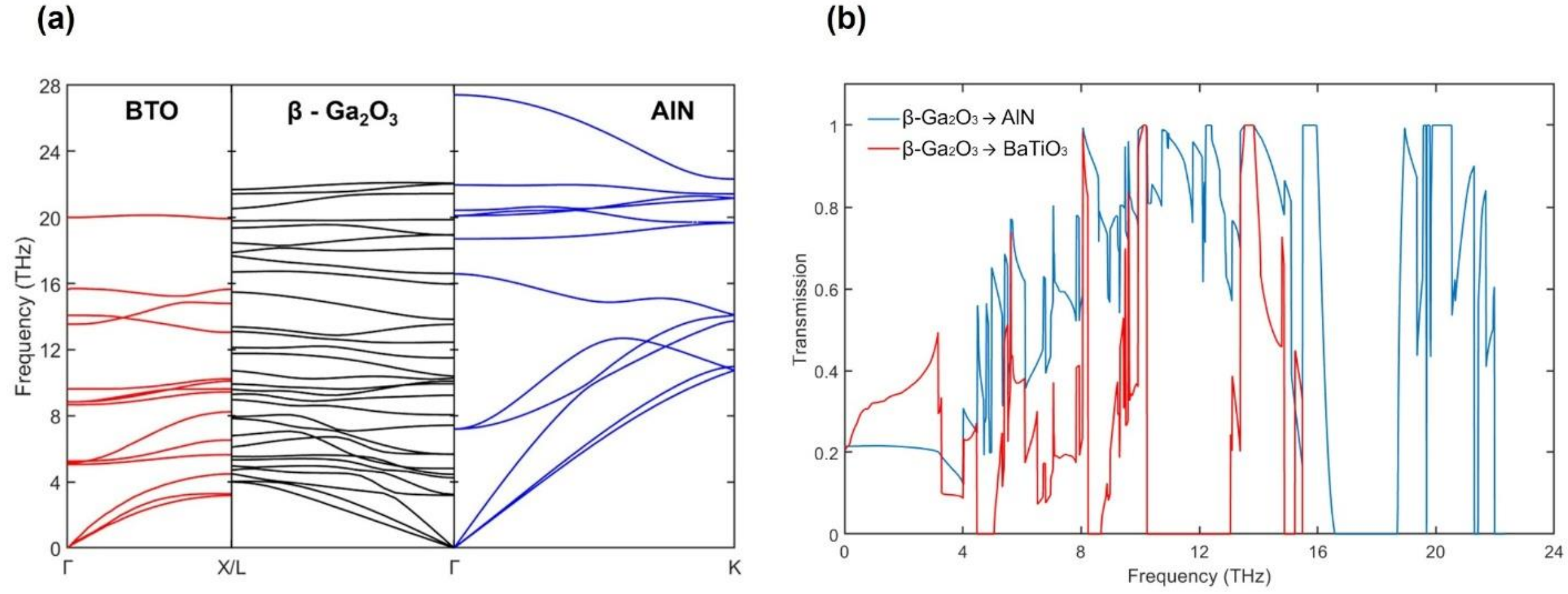

Figure 7. (a) The phonon dispersion relations of $BaTiO_3$, β-$Ga_2O_3$, and AlN (b) The transmission coefficients of β-$Ga_2O_3$ to $BaTiO_3$ and β-$Ga_2O_3$ to AlN from DMM.

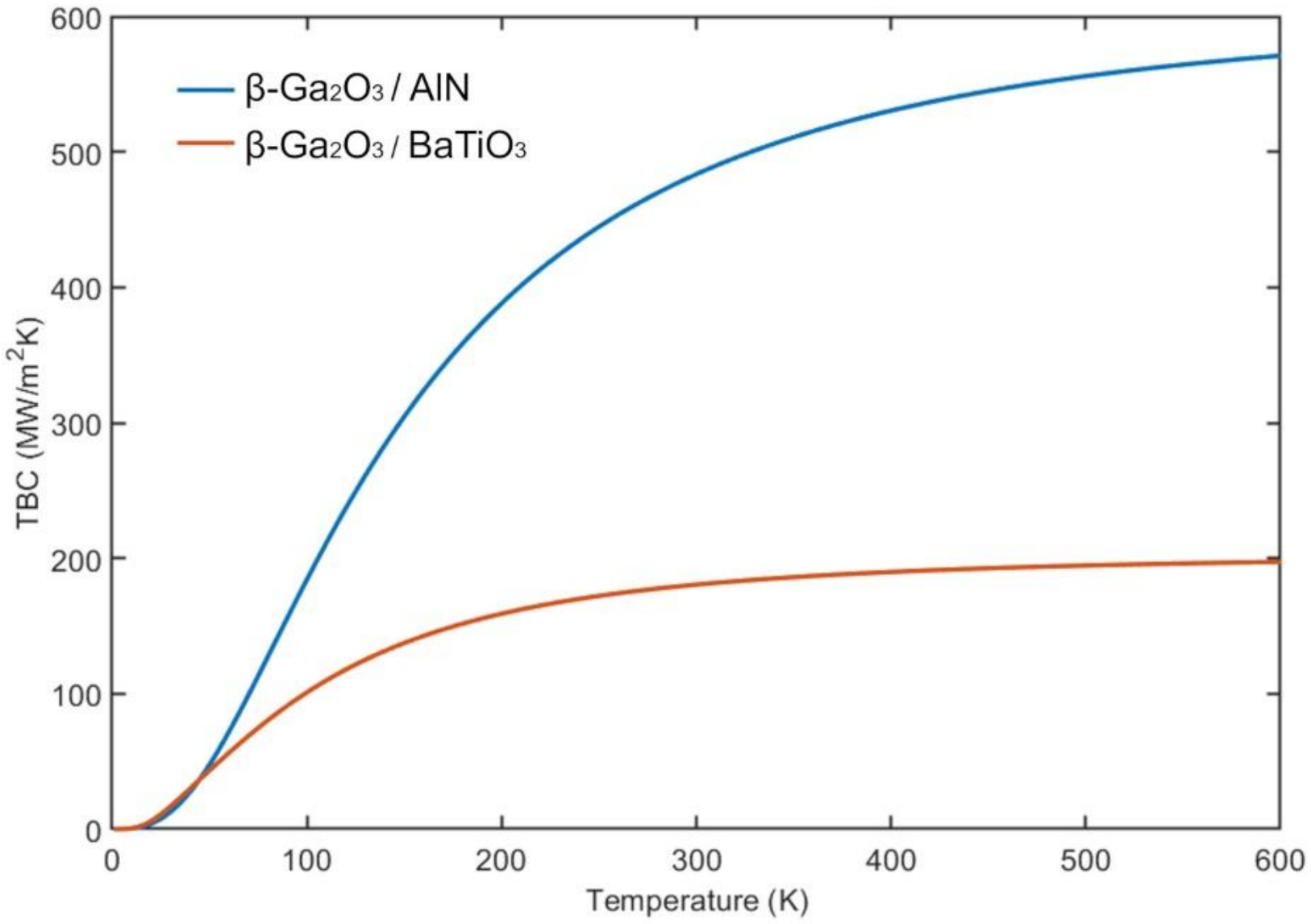

Figure 8. TBC at β-$Ga_2O_3$/$BaTiO_3$ and β-$Ga_2O_3$/AlN interfaces as a function of temperature.

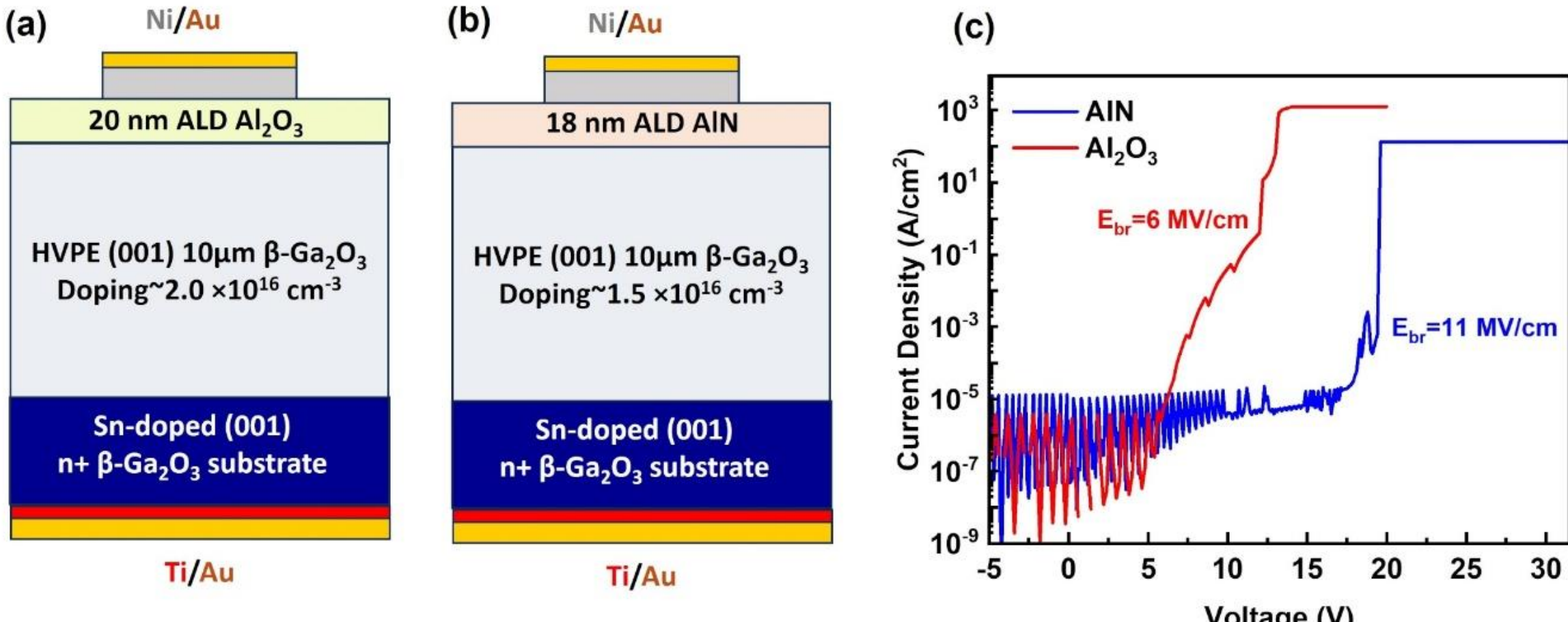

Figure 9. Fabricated vertical β-$Ga_2O_3$ MIS diode structures with (a) ALD $Al_2O_3$, (b) ALD AlN, and (c) forward breakdown characteristics of the MIS diodes demonstrating significantly higher breakdown field exhibited by AlN (~11 MV/cm) compared to that of $Al_2O_3$ (~6 MV/cm).